\documentclass[10pt,leqno]{amsart}
\usepackage{graphicx}
\usepackage{indentfirst,csquotes}

\usepackage{graphicx}
\usepackage{indentfirst,csquotes}
\usepackage{amssymb,amsthm,amsmath}
\usepackage{xcolor,paralist,hyperref,fancyhdr,etoolbox} 
\usepackage{braket}
\usepackage{subcaption} 
\usepackage{booktabs} 
\usepackage{tikz}
\usepackage{pgfplots}
\usepackage{lmodern}
\usepackage{import}
\usepackage{algorithm}
\usepackage{algpseudocode}
\usetikzlibrary{automata, positioning}
\usetikzlibrary{shapes.geometric, arrows, positioning}

\everymath=\expandafter{\the\everymath\displaystyle}

\tikzstyle{block} = [rectangle, draw, text centered, rounded corners, minimum height=2em]
\tikzstyle{line} = [draw, -stealth, thick]
\tikzstyle{cloud} = [ellipse, draw, text centered, minimum height=2em, thick]
\tikzstyle{dashedcloud} = [ellipse, draw, dashed, text centered, minimum height=2em, thick]
\usetikzlibrary{positioning, arrows.meta, shapes.geometric, decorations.pathreplacing}

\tikzstyle{startstop} = [rectangle, rounded corners, minimum width=1.5cm, minimum height=0.5cm,text centered, draw=black, fill=red!30]
\tikzstyle{io} = [trapezium, trapezium left angle=70, trapezium right angle=110, minimum width=1cm, minimum height=0.5cm, text centered, draw=black, fill=blue!30]

\tikzstyle{process} = [rectangle, minimum width=3cm, minimum height=0.5cm, text centered, draw=black, fill=orange!30]
\tikzstyle{decision} = [diamond, minimum width=0.5cm, minimum height=0.1cm, text centered, draw=black, fill=green!30]
\tikzstyle{process2} = [rectangle, minimum width=1cm, minimum height=0.5cm, text centered, draw=black, fill=orange!30]
\tikzstyle{arrow} = [thick,->,>=stealth]
\usetikzlibrary{shapes,shadows,arrows,positioning,angles,quotes}
\tikzset{My Arrow Style/.style={single arrow, fill=black!15, anchor=base, align=center,text width=2.3cm}}
\tikzstyle{arrow} = [thick,->,>=stealth]
\usetikzlibrary{calc,trees,positioning,arrows,fit,shapes,calc}

\newtheorem{theorem}{Theorem}
\newtheorem{definition}[theorem]{Definition}

\hypersetup{
  colorlinks=true,
  linkcolor=blue,    
  citecolor=blue,    
  filecolor=blue,    
  urlcolor=blue      
}

\begin{document}

\title{Formal Verification of Noisy Quantum Reinforcement Learning Policies}
\author[Dennis Gross]{Dennis Gross}
\maketitle


\begin{abstract}
\emph{Quantum reinforcement learning (QRL)} aims to leverage quantum phenomena to develop sequential decision-making policies that achieve task objectives more effectively than their classical counterparts.
Unlike reinforcement learning policies implemented on classical hardware, QRL policies are subject to additional uncertainty from quantum measurements and hardware imperfections.
These uncertainties, including bit-flip, phase-flip, and depolarizing noise, can alter agent behavior and lead to unsafe policy behaviour.
However, there do not exist approaches that provide systematic methods for verifying \emph{exactly} whether trained QRL policies satisfy safety requirements under specific quantum noise conditions.
We present \emph{QVerifier}, a formal verification method that uses \emph{probabilistic model checking} to rigorously analyze trained QRL policies, with and without modeled quantum noise, to verify satisfaction or violation with safety properties.
Our method incrementally constructs a formal model of the policy–environment interaction by expanding all states reachable under the policy and assigning their transition probabilities using the model's dynamics.
Quantum measurement uncertainty and optional quantum hardware noise are incorporated directly into these probabilities.
When the policy was trained under matching noise conditions, this formal model is exact; when trained on physical hardware, it constitutes an idealized approximation, as unknown hardware noise prevents exact modeling.
The resulting model is then verified using the \emph{Storm} model checker to assess satisfaction of the safety properties.
Experiments across multiple QRL environments show that our approach precisely quantifies how different models of quantum noise affect safety guarantees. The results reveal not only how specific noise models degrade policy performance, but also cases where certain noise can improve it.
By enabling rigorous safety verification before deployment, \emph{QVerifier} addresses a critical need: because access to quantum hardware is prohibitively expensive, pre-deployment verification is essential for any safety-critical use of quantum reinforcement learning.
It targets a potential sweet spot between classical and quantum computation, where trained QRL policies could still be modeled classically for probabilistic model checking.
\end{abstract}


\section{Introduction}
As \emph{quantum computing}~\cite{nielsen2010quantum} transitions from theory to commercial reality~\cite{macquarrie2020emerging,18166,18276}, it opens new opportunities for advancing \emph{machine learning}~\cite{DBLP:journals/ncs/CerezoVHCC22,bano2025envisioning,18277}. \emph{Quantum reinforcement learning (QRL)} combines \emph{reinforcement learning (RL)} with quantum computing to train sequential decision-making policies that may achieve task objectives more effectively than their classical counterparts by leveraging quantum phenomena such as superposition and entanglement~\cite{jerbi2021quantum,chen2024introduction,meyer2022survey,skolik2022quantum,DBLP:journals/corr/abs-2505-11862,jerbi2021parametrized}.

In general, an RL agent aims to learn a near-optimal policy to achieve a fixed objective by taking actions and receiving feedback in the form of rewards and state observations from the environment~\cite{sutton2018reinforcement}.
A policy is \emph{memoryless} if it depends only on the current state, and it can be \emph{stochastic}, assigning action probabilities from which the agent samples its next action.

In QRL, policies are encoded by \emph{variational quantum circuits (VQCs)}~\cite{skolik2022quantum,DBLP:journals/qmi/SequeiraSB23,DBLP:conf/qce/SequeiraSB24}, which are parameterized circuits optimized during training.
A VQC takes a quantum encoding of the environment state and produces a quantum state, from which action probabilities are estimated via \emph{repeated measurements (shots)}.
Each shot measures the circuit on the \emph{same encoded input}, yielding a single classical outcome by projecting the quantum state onto one of its possible results; many such shots are needed to approximate the underlying action distribution.
VQCs can be executed on quantum simulators or quantum hardware~\cite{wu2021application}, though simulating larger circuits becomes more and more difficult~\cite{DBLP:journals/corr/abs-2410-12660}.

As research in QRL progresses into \emph{safety-critical systems}~\cite{DBLP:conf/icassp/KimCPK25,niraula2021quantum,lv2025supply,dunn2025q}, it is crucial to ensure that these QRL policies are developed with safety in mind~\cite{bano2025envisioning,skolik2023robustness}.

However, trained policies may exhibit \emph{unsafe behavior}~\cite{DBLP:conf/setta/GrossJJP22}, such as collisions~\cite{DBLP:journals/access/BanL24,DBLP:journals/qmi/SinhaMK25}, since rewards does not capture complex safety requirements~\cite{DBLP:journals/aamas/VamplewSKRRRHHM22}.
In QRL, additional uncertainty from stochastic measurements and device noise~\cite{DBLP:conf/hais/EscuderoAGB23,DBLP:journals/quantum/ArochKK24} on costly quantum computing hardware~\cite{DBLP:journals/sqj/MoguelRVBGM22} can further distort action selection.

To resolve the safety issues mentioned above, formal verification methods like \emph{probabilistic model checking}~\cite{DBLP:conf/setta/GrossJJP22} have been proposed to reason about the safety of RL on classical hardware~\cite{yuwangPCTL,DBLP:conf/formats/HasanbeigKA20,DBLP:conf/atva/BrazdilCCFKKPU14,DBLP:conf/tacas/HahnPSSTW19} and the correctness of quantum systems~\cite{gay2005probabilistic,baltazar2008quantum}.
Model checking is not limited by properties that rewards can express but supports a broader range of properties that can be expressed by \emph{probabilistic computation tree logic (PCTL)} \cite{hansson1994logic}, a temporal logic designed to express probabilistic branching-time properties.
At its core, probabilistic model checking uses mathematical models to verify a system's correctness with respect to a given safety property~\cite{baier2008principles}.

However, in the context of QRL policy verification, probabilistic model checking has not yet been applied, and there are limited tools for systematic comparisons of QRL policies~\cite{kaldari2025quantum,bowles2024better,kruse2025benchmarking}.

\emph{In this paper}, we introduce \emph{QVerifier}, a method for model-checking trained QRL policies against safety properties while explicitly accounting for quantum uncertainty arising from measurements and modeled quantum hardware noise.
Given a trained QRL policy, an assumed quantum noise model, and a formal environment model, we incrementally build a formal model of the combined policy–noise–environment system, exploring only the states reachable under the policy.
When the policy was trained on a simulator, this model is exact; when trained on physical hardware, it constitutes an idealized approximation, as the true noise experienced during training and deployment may differ from the assumed model.
After the building process, we use the probabilistic model checker \emph{Storm}~\cite{DBLP:journals/sttt/HenselJKQV22} to verify whether the resulting formal model satisfies the desired safety property.

\emph{Our experiments} demonstrate that probabilistic model checking enables rigorous verification of trained QRL policies with and without noise with respect to safety properties, before deploying them on costly quantum hardware.
We verify six policies (three QRL and three classical RL baselines) across three environments (Frozen Lake, Ski, and Freeway) under four quantum noise models: bit-flip, phase-flip, depolarizing, and amplitude damping that are applied independently at each quantum circuit gate, though \emph{QVerifier} is not limited to these noise models.
Our results reveal that noise effects are policy- and task-dependent: while bit-flip and depolarizing noise consistently degrade performance, we observe that low-level amplitude-damping noise can sometimes improve QRL policy performance, with the Ski environment showing a 27\% improvement over the classical RL baseline.
These findings demonstrate that pre-deployment verification not only identifies safety violations but also uncovers beneficial quantum noise regimes.

Our \textbf{main contributions} are a method for formally verifying trained QRL policy behaviors in modeled environments, with and without quantum noise models, and an exact comparison approach for QRL and classical RL performance.
By enabling rigorous safety verification before deployment, \emph{QVerifier} addresses a critical need: access to quantum hardware is prohibitively expensive, making pre-deployment verification essential for safety-critical applications.
\emph{QVerifier} targets a potential sweet spot between classical and quantum computation, where trained QRL policies could still be modeled classically for probabilistic model checking.

\section{Related Work}
Research at the intersection of quantum computing, machine learning, and software engineering has grown rapidly in recent years~\cite{bano2025envisioning}.  
A key challenge across these areas is dealing with quantum noise, which affects both quantum circuits and the software built on top of them.  
Noise-aware quantum software engineering, therefore, forms an active research direction~\cite{16888}.

\subsection*{Verification of Quantum Systems}
Probabilistic model checking has long been used to verify the correctness of quantum systems, such as quantum communication protocols~\cite{gay2005probabilistic} or quantum key distribution~\cite{baltazar2008quantum}.  
More recently, verification methods for quantum machine-learning classifier models have emerged.  
For example, Guan et al.\ propose a robustness verification method for quantum classifiers~\cite{DBLP:conf/cav/GuanFY21}, and \emph{VeriQR} provides automated robustness checking for such classifiers~\cite{DBLP:conf/fm/LinGFYS24}.  
However, none of these works apply probabilistic model checking to the verification of \emph{QRL policies}.

\subsection*{Verification of Classical RL Policies}
In RL on classical hardware, several works study the verification of trained RL policies~\cite{DBLP:conf/sigcomm/EliyahuKKS21,DBLP:conf/sigcomm/KazakBKS19,pmlr-v161-corsi21a,DBLP:journals/corr/DragerFK0U15,DBLP:conf/pldi/ZhuXMJ19,DBLP:conf/seke/JinWZ22}.  
A particularly relevant line of work, for this paper, uses probabilistic model checking to verify deep RL policies~\cite{DBLP:conf/setta/GrossJJP22,DBLP:conf/icaart/GrossS24,DBLP:journals/access/KwonK25}.  
\emph{COOL-MC}~\cite{DBLP:conf/setta/GrossJJP22} introduced incremental model construction for RL policies and safety verification with the \emph{Storm} model checker.  
Our work builds directly on this idea but extends it to handle QRL policies, including quantum measurements and hardware noise.

\subsection*{Noise and Robustness in RL and QRL}
Classical noise has also been studied in the context of classical RL, where perturbations typically affect the agent's state observations rather than the policy's inner workings~\cite{DBLP:conf/icaart/GrossS0023,DBLP:conf/aips/GrossS0023}.  
In contrast, QRL naturally introduces additional uncertainty from quantum-state measurements and noise arising from error-prone quantum circuit gates within the quantum circuit.

For improving robustness of QRL during training, \emph{RegQPG} adds Lipschitz regularization to the policy gradient objective~\cite{DBLP:journals/corr/abs-2410-21117}.  
Unlike this training-time approach, we perform \emph{post-training verification} to check whether a trained QRL policy satisfies a safety property under explicit hardware noise models (bit-flip, phase-flip, depolarizing, amplitude-damping).

Several works analyze how quantum noise affects QRL performance.  
Skolik et al.\ study realistic noise sources, including shot noise, coherent Gaussian noise, and decoherence, and show that their effects vary strongly with both the type and the magnitude of the noise~\cite{skolik2023robustness}.  
These insights motivate our goal of \emph{exactly} verifying trained QRL policies \emph{under specific noise models}, rather than studying robustness in aggregate.

\subsection*{Other Related Directions}
Some research modifies trained RL policy architecture by disrupting parts of it during verification~\cite{DBLP:conf/esann/GrossS24}, but this differs from having QVC gate-level noise.
Closely related to verification is testing, with extensive work on both testing quantum systems and using quantum computing for testing~\cite{18308,muqeet2024assessing,wang2025bqtmizer,oldfield2025faster,mendiluze2025quantum,wang2025quantum,araujo2025using}.

\medskip
\noindent
\textbf{Summary.}  
While verification techniques exist for quantum systems, quantum classifiers, and classical RL policies, no prior work uses probabilistic model checking to verify \emph{memoryless QRL policies} under different quantum noise models.  
Our work fills this gap by introducing a QRL-specific model-checking method.

\section{Background}
First, we outline how sequential decision-making tasks can be formally represented as \emph{Markov decision processes (MDPs)}, and how memoryless stochastic policies and their properties can be verified with this method.
Second, we introduce quantum circuits and describe how to model relevant sources of quantum noise.
Finally, we present quantum REINFORCE, a QRL method for training sequential decision-making policies.

\subsection{Probabilistic Systems}
A \textit{probability distribution} over a set $X$ is a function $\mu \colon X \rightarrow [0,1]$ with $\sum_{x \in X} \mu(x) = 1$. The set of all distributions on $X$ is $Distr(X)$.

\begin{definition}[MDP]\label{def:mdp}
A \emph{MDP} is a tuple $M = (S,s_0,Act,Tr, rew, AP,L)$
where $S$ is a finite, nonempty set of states; $s_0 \in S$ is an initial state; $Act$ is a finite set of actions; $Tr\colon S \times Act \rightarrow Distr(S)$ is a partial probability transition function and $Tr(s,a,s')$ denotes the probability of transitioning from state $s$ to state $s'$ when action $a$ is taken;
$rew \colon S \times Act \rightarrow \mathbb{R}$ is a reward~function;
$AP$ is a set of atomic propositions;
$L \colon  S \rightarrow 2^{AP}$ is a labeling~function.
\end{definition}

We represent each state $s \in S$ as a vector of $d$ integer features $(f_1, \dots, f_d)$, where $f_i \in \mathbb{Z}$.
The available actions in $s \in S$ are $Act(s) = \{a \in Act \mid Tr(s,a) \neq \bot\}$ where $Tr(s, a) \neq \bot$ is defined as action $a$ at state $s$ does not have a transition (action $a$ is not available in state $s$).
In our setting, we assume that all actions are available at all states.


\begin{definition}[Memoryless Stochastic Policy]
A \emph{memoryless stochastic policy} $\pi$ for an MDP $M$ is a function
\[
\pi \colon S \rightarrow Distr(Act),
\]
i.e., for each state $s \in S$, $\pi(s)$ is a probability distribution over actions. Equivalently, this can be expressed as
\[
\pi(a \mid s) = \pi(s)(a) \in [0,1], \quad \sum_{a \in Act} \pi(a \mid s) = 1,
\]
where $\pi(a \mid s)$ denotes the probability of selecting action $a$ when in state $s$.
\end{definition}


Applying a memoryless stochastic policy $\pi$ to an MDP $M$ yields an \emph{induced discrete-time Markov chain (DTMC)} whose transition probabilities combine the MDP's probabilistic dynamics and the policy's stochastic choice of actions.

\begin{definition}[Induced DTMC]\label{def:induced-dtmc}
Given an MDP $M = (S,s_0,Act,Tr, rew, AP,L)$ and a stochastic policy $\pi$, the \emph{induced DTMC} is $M^\pi = (S, s_0, P^\pi, AP, L)$ where the transition probability function $P^\pi \colon S \times S \rightarrow [0,1]$ is defined by marginalizing over actions
\[
P^\pi(s,s') = \sum_{a \in Act(s)} \pi(a \mid s) \cdot Tr(s,a,s').
\]
\end{definition}

The transition probability $P^\pi(s,s')$ represents the total probability of moving from state $s$ to state $s'$ by summing over all actions weighted by their selection probabilities under policy~$\pi$.

\subsection{Probabilistic Model Checking}

Storm~\cite{DBLP:journals/sttt/HenselJKQV22} is a model checker. 
It enables the verification of properties in induced DTMCs, with reachability properties being among the most fundamental.
These properties assess the probability of a system reaching a particular state.
For example, one might ask, ``Is the probability of the system reaching an unsafe state less than 0.1?''
A property can be either \emph{satisfied} or \emph{violated}.

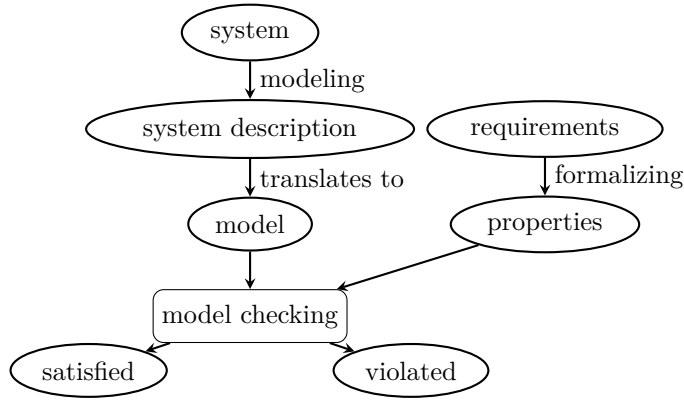
\begin{figure}[htbp]
    \centering
    \begin{tikzpicture}[node distance=0.5cm, auto]
    \node [cloud] (system) {system};
    \node [cloud, below=0.5cm of system] (systemdesc) {system description};
    \node [cloud, below=0.5cm of systemdesc] (model) {model};
    \node [cloud, right=0.15cm of systemdesc] (requirements) {requirements};
    \node [cloud, below=0.5cm of requirements] (properties) {properties};
    \node [block, below=0.5cm of model] (modelchecking) {model checking};
    \node [cloud, below left=0.15cm of modelchecking] (satisfied) {satisfied};
    \node [cloud, below right=0.15cm of modelchecking] (violated) {violated};

    \path [line] (system) -- node[right] {modeling} (systemdesc);
    \path [line] (systemdesc) -- node[right] {translates to} (model);
    \path [line] (requirements) -- node[right] {formalizing} (properties);
    \path [line] (model) -- (modelchecking);
    \path [line] (properties) -- (modelchecking);
    \path [line] (modelchecking) -- (satisfied);
    \path [line] (modelchecking) -- (violated);
    \end{tikzpicture}
    \caption{General model checking workflow~\cite{DBLP:journals/sttt/HenselJKQV22}. First, the system needs to be formally modeled, for instance, via PRISM. Then, the requirements are formalized, for instance, via PCTL. Eventually, both are inputted into the model checker, like Storm, which verifies the property.}
    \label{fig:model_checking}
\end{figure}

The \emph{general workflow} for model checking with Storm is as follows (see also Figure~\ref{fig:model_checking}):
First, the system, in our setting, an induced DTMC, is modeled using a language such as PRISM~\cite{prism_manual}.
Next, a property is formalized based on the system's requirements.
Using these inputs, the model checker Storm verifies whether the formalized property holds or fails within the model.

In probabilistic model checking, there is no universal ``one-size-fits-all'' solution~\cite{DBLP:journals/sttt/HenselJKQV22}.
The most suitable tools and techniques depend on the specific input model and properties being analyzed.
During model checking, Storm can proceed ``on the fly'', exploring only the parts of the DTMC most relevant to the formal verification.

Properties verifiable via Storm include temporal logic formulas for DTMCs, defined on paths using PCTL~\cite{DBLP:journals/fac/HanssonJ94}, a branching-time logic.

Although it is not formally allowed in PCTL, Storm can still be used to request the probability of fulfilling a path formula from each state.
Rather than simply checking whether certain PCTL properties, such as $P_{=1}(\text{F } \textit{collision})$, are satisfied, which would indicate that the system reaches the $collision$ state with probability 1, we can query Storm to compute the actual probability value, denoted as $P(\text{F } \textit{collision})$, which in this case equals 1.

In this paper, we consider the \emph{eventually} path property $\text{F }$, which states that a property holds at some future state along a path. Its syntax is
\[
P_{\sim p} (\text{F } \varphi)
\]
meaning that the probability of reaching a state where $\varphi$ holds, along a path starting from the initial state, satisfies the bound $\sim p$, where $\sim$ is a comparison operator such as $<$, $\leq$, $\geq$, or $>$, and $p$ is a probability threshold.

We also consider the \emph{until} path property $U$, which is more expressive than simple reachability. The until property states that a condition $\varphi$ holds along a path until a state satisfying $\psi$ is reached. Its syntax is
\[
P_{\sim p} (\varphi \text{ U } \psi)
\]
meaning that the probability of reaching a state where $\psi$ holds, while $\varphi$ holds continuously along the path until that point, satisfies the bound $\sim p$.

\subsection{Quantum Computing}

A quantum state $\ket{\psi}$ of a system is represented by a unit vector in a complex Hilbert space $\mathcal{H}$. This is  $\mathcal{H}= \mathbb{C}^2$ for a single qubit system~\cite{nielsen2010quantum}.
In Dirac notation, $\ket{\psi}$ is called a \emph{ket} and represents a column vector, while its dual $\bra{\psi}$ is called a \emph{bra} and represents the corresponding conjugate transpose (row vector).
The inner product between two states is written as $\langle\phi|\psi\rangle$, and the outer product $\ket{\phi}\bra{\psi}$ forms a matrix.

More generally, quantum systems can be represented by a \emph{density matrix} $\rho = \ket{\psi}\bra{\psi}$.
Density matrices are essential for describing mixed states (statistical mixtures of pure states) and systems affected by noise, which cannot be represented by a single state vector.
For a general mixed state, we have
\[
\rho = \sum_i p_i \ket{\psi_i}\bra{\psi_i},
\]
where $\{p_i\}$ are classical probabilities ($p_i \ge 0$, $\sum_i p_i = 1$).

Every valid density matrix satisfies three key properties: it is \emph{Hermitian}, meaning it is equal to its conjugate transpose and thus has real eigenvalues; \emph{positive semidefinite}, ensuring that all probabilities are non-negative; and has \emph{unit trace}, ensuring that probabilities of all measurement outcomes sum to one.

\subsubsection{Entanglement}
A multi-qubit quantum state is called \emph{entangled} if it cannot be written as a tensor product of individual qubit states.
For a two-qubit system, a state $\ket{\psi} \in \mathcal{H}_A \otimes \mathcal{H}_B$ is \emph{separable} if it can be expressed as
\[
\ket{\psi} = \ket{\phi}_A \otimes \ket{\chi}_B,
\]
and \emph{entangled} otherwise.
Entangled states produce measurement correlations that are impossible to replicate with any classical system~\cite{rosset2013entangled}.

\subsubsection{Amplitude encoding}
A common technique for encoding classical data vectors into quantum states is \emph{amplitude encoding}~\cite{kaldari2025quantum}.
Given a classical state $s \in S$ with $d$ components, amplitude encoding prepares a quantum state
\[
\ket{\psi} = \sum_{i=0}^{d-1} \tilde{s}_i \ket{e_i},
\]
where $\{\ket{e_i}\}_{i=0}^{d-1}$ is an orthonormal basis for $\mathcal{H} = \mathbb{C}^d$, and $\tilde{s}_i = s_i / \|s\|$ are the normalized amplitudes, ensuring $\sum_i |\tilde{s}_i|^2 = 1$.

\subsubsection{Quantum channels}

Quantum evolution is described by a \emph{quantum channel}, a linear, completely positive, and trace-preserving map $\mathcal{E} : \rho \mapsto \rho'$.
Every quantum channel admits a \emph{Kraus representation}
\[
\mathcal{E}(\rho) = \sum_i K_i \rho K_i^\dagger,
\]
where $\{K_i\}$ are Kraus operators satisfying $\sum_i K_i^\dagger K_i = I$.
The \emph{completely positive} property ensures that the map remains physical when applied to subsystems, while \emph{trace-preserving} guarantees that probabilities sum to one.

\medskip
\noindent\emph{Unitary evolution.}
The special case of \emph{noiseless} quantum evolution corresponds to a single Kraus operator $K = U$, where $U$ is a \emph{unitary operator} satisfying $U^\dagger U = UU^\dagger = I$.
The channel becomes
\[
\mathcal{E}(\rho) = U\rho U^\dagger.
\]
Unitary operations are \emph{reversible} (applying $U^\dagger$ inverts the operation) and preserve quantum~coherence.

\medskip
\noindent\emph{Pauli operators.}
The Pauli operators $\{X, Y, Z\}$ are fundamental single-qubit operators:
\[
X = \begin{pmatrix} 0 & 1 \\ 1 & 0 \end{pmatrix}, \quad
Y = \begin{pmatrix} 0 & -i \\ i & 0 \end{pmatrix}, \quad
Z = \begin{pmatrix} 1 & 0 \\ 0 & -1 \end{pmatrix}.
\]
These operators are both Hermitian ($P^\dagger = P$) and unitary ($P^\dagger P = I$).

\subsubsection{Noisy channels}
These channels arise when quantum systems interact with their environment and require multiple Kraus operators.
Each $K_i$ corresponds to a possible outcome of an interaction with the environment or an error process.
Noisy channels with multiple Kraus operators are \emph{irreversible}.
Different physical noise processes correspond to specific choices of Kraus operators.

\begin{itemize}
    \item \emph{Bit-flip noise:}
    \[
    \mathcal{E}_{\text{BF}}(\rho) = (1-p)\rho + p\, X\rho X,
    \]
    where $p$ is the bit-flip probability.
    
    \item \emph{Phase-flip noise:}
    \[
    \mathcal{E}_{\text{PF}}(\rho) = (1-p)\rho + p\, Z\rho Z.
    \]
    
    \item \emph{Depolarizing noise:}
    \[
    \mathcal{E}_{\text{DP}}(\rho) = (1-p)\rho + \frac{p}{3}(X\rho X + Y\rho Y + Z\rho Z).
    \]
    
    \item \emph{Amplitude damping noise:}
    \[
    \mathcal{E}_{\text{AD}}(\rho) = E_0 \rho E_0^\dagger + E_1 \rho E_1^\dagger,
    \]
    with Kraus operators
    \[
    E_0 = 
    \begin{pmatrix}
    1 & 0 \\
    0 & \sqrt{1-\gamma}
    \end{pmatrix},
    \quad
    E_1 = 
    \begin{pmatrix}
    0 & \sqrt{\gamma} \\
    0 & 0
    \end{pmatrix},
    \]
    where $\gamma$ denotes the damping rate, representing the probability of energy loss from the excited to the ground state.
\end{itemize}

\subsubsection{Computing measurement probabilities analytically}
When modeling a quantum circuit analytically (e.g., using exact matrix 
calculations and noise models), the measurement probabilities can be 
directly extracted from the final density matrix $\rho_{\text{final}}$.

For computational basis measurements, the probability of measuring 
outcome $x$ is simply given by the corresponding diagonal element
\[
p_x = \bra{x}\rho_{\text{final}}\ket{x} = (\rho_{\text{final}})_{xx}.
\]

For a single qubit, this means
\[
p_0 = \rho_{00}, \qquad p_1 = \rho_{11}.
\]

For an $n$-qubit system, the density matrix is $2^n \times 2^n$, and 
the probability of measuring bitstring $x \in \{0,1\}^n$ is the diagonal 
element at the corresponding index.

\subsubsection{Multi-shot experiments}
On quantum hardware, the outcome of a quantum measurement is inherently probabilistic: each circuit execution (or \emph{shot}) yields a single bitstring sampled from the probability distribution $p_x$.
To estimate these probabilities, one repeats the circuit many times (a process known as performing a \emph{multi-shot experiment}).
After $N$ shots, the empirical frequencies $\hat{p}_x = n_x / N$ (where $n_x$ is the count of outcome $x$) approximate the true probabilities~$p_x$.

\subsection{Quantum REINFORCE}

\begin{figure}[htbp]
\centering
\scalebox{1}{
\begin{tikzpicture}[
    node distance=2.5cm,
    every node/.style={},
    process/.style={
        rectangle,
        rounded corners,
        draw=black,
        thick,
        minimum width=2.5cm,
        minimum height=1cm,
        align=center
    },
    arrow/.style={
        ->,
        thick,
        >=stealth
    }
]
\node (agent) [process] {Policy};
\node (env) [process, right of=agent, xshift=2.5cm] {Environment};

\draw [arrow] (agent) -- node[above, font=\small] {Action} (env);
\draw [arrow] (env) .. controls +(0,-1.2) and +(0,-1.2) .. node[below, font=\small] {New State, Reward} (agent);
\end{tikzpicture}
}
\caption{An RL agent interacts with an environment. The agent sends an action and receives the resulting state and reward, closing the feedback loop. The interaction ends when a terminal state is reached.}
\label{fig:rl}
\end{figure}
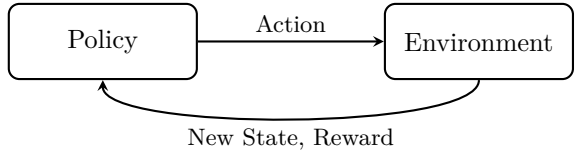

\begin{figure}[]
    \centering
    \includegraphics[width=\columnwidth]{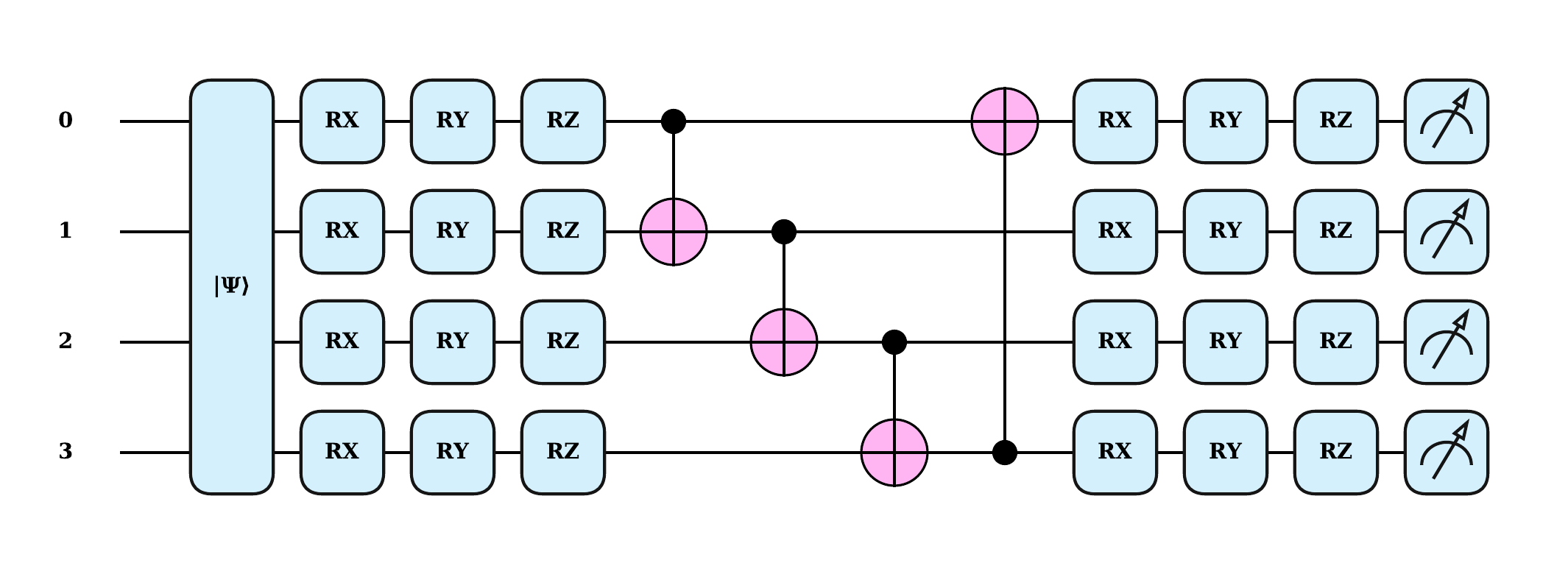}
    \caption{An example of a variational quantum circuit architecture for a quantum REINFORCE agent. 
    The circuit operates on four qubits and has two variational layers with 24 trainable parameters (gates with name R*). 
    State preparation encodes the input via amplitude encoding to the qubits. Layer 0 contains rotation gates 
    (RX, RY, RZ) and CNOT entanglement in ring topology. Layer 1 contains only rotations. 
    Pauli-Z measurements produce expectation values that are classically post-processed into 
    action probabilities via a softmax function.}
    \label{fig:quantum_circuit}
\end{figure}

REINFORCE~\cite{williams1992simple} is a policy gradient RL algorithm that trains a parameterized policy $\pi_\theta$ by directly optimizing the expected cumulative reward (see Figure~\ref{fig:rl}).
Given the reward function $rew$ from Definition~\ref{def:mdp}, the \emph{return} $G_t$ from time step $t$ is the discounted sum of rewards along a trajectory
\[
G_t = \sum_{k=t}^T \gamma^{k-t} \, rew(s_k, a_k),
\]
where $T$ is the episode length and $\gamma \in [0,1]$ is the discount factor.
The policy gradient theorem~gives
\[
\nabla_\theta J(\theta) = \mathbb{E}_{\pi_\theta}\left[\sum_{t=0}^T \nabla_\theta \log \pi_\theta(a_t \mid s_t) \cdot G_t\right],
\]
where $J(\theta)$ is the expected return and the expectation is over trajectories sampled from policy $\pi_\theta$.
Parameters are updated via gradient ascent using sampled trajectories.

In \emph{Quantum REINFORCE}, the policy $\pi_\theta$ is implemented as a parameterized quantum circuit (see Figure~\ref{fig:quantum_circuit}), where $\theta$ represents quantum gate parameters~\cite{sequeira2023policy}.
For state $s$, the circuit prepares a density matrix $\rho_\theta(s)$.
Performing a computational basis measurement yields action~probabilities
\[
\pi_\theta(a \mid s) = \bra{a}\rho_\theta(s)\ket{a} = \mathrm{Tr}\!\left(\ket{a}\bra{a} \, \rho_\theta(s)\right).
\]
The gradient $\nabla_\theta \log \pi_\theta(a \mid s)$ is computed using parameter-shift rules~\cite{mitarai2018quantum}, and parameters are updated following standard REINFORCE.

Given a policy $\pi$ with parameters $\theta$ and a noise channel $\mathcal{E}$, action selection for state $s$ proceeds as follows.
The state $s$ is encoded into an initial density matrix $\rho_0(s)$, which is then transformed by the trained unitary circuit $U_{\theta}$ to produce $\rho = U_{\theta}\rho_0(s)U_{\theta}^\dagger$.
The noise channel is applied to yield the noisy density matrix $\rho' = \mathcal{E}(\rho)$, and measurement gives action probabilities
\[
\pi_{\theta}^{\mathcal{E}}(a \mid s) = \bra{a}\mathcal{E}\!\left(U_{\theta}\rho_0(s)U_{\theta}^\dagger\right)\ket{a}.
\]

\section{Methodolodgy}
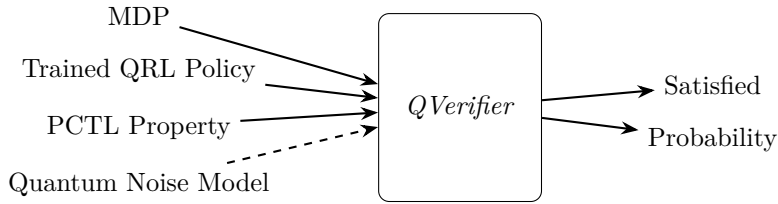
\begin{figure}
    \centering
\begin{tikzpicture}[>=Stealth, node distance=0.25cm]
    \node [] (input1) {MDP};
    \node [below=0.2cm of input1] (input2) {Trained QRL Policy};
    \node [below=0.2cm of input2] (input3) {PCTL Property};
    \node [below=0.2cm of input3] (input4) {Quantum Noise Model};
    
    \node [right=1.5cm of input2, draw, rounded corners, inner sep=10pt, minimum height=2.5cm, yshift=-0.5cm] (qverifier) {\emph{QVerifier}};
    
    \node [right=1.5cm of qverifier, yshift=0.3cm] (output1) {Satisfied};
    \node [below=0.2cm of output1] (output2) {Probability};
    
    \draw[->, thick] (input1) -- (qverifier);
    \draw[->, thick] (input2) -- (qverifier);
    \draw[->, thick] (input3) -- (qverifier);
    \draw[->, thick, dashed] (input4) -- (qverifier);
    
    \draw[->, thick] (qverifier) -- (output1);
    \draw[->, thick] (qverifier) -- (output2);
\end{tikzpicture}
    \caption{Given an MDP, a trained QRL policy, a PCTL property, and, optionally, quantum noise models as inputs, the method outputs whether the safety property is satisfied and its exact probability.}
    \label{fig:workflow}
\end{figure}

As illustrated in Figure~\ref{fig:workflow}, our method, \emph{QVerifier}, takes as input: (i) an MDP representing the environment dynamics, (ii) a trained QRL policy, (iii) a PCTL property specifying the desired safety requirement, and (iv) an optional quantum noise model capturing hardware imperfections.
\emph{QVerifier} verifies whether the policy satisfies or violates the property and quantifies the probability for it.
Our approach consists of two internal stages (see also Algorithm~\ref{alg:qverifier}): (i) constructing the induced DTMC, and (ii) performing probabilistic model checking of the built formal model.

\subsection{Induced DTMC Construction}
Given an MDP $M$ and a memoryless stochastic policy $\pi$, where $\pi(a \mid s)$ denotes the probability of selecting action $a$ in state $s$, we construct an induced DTMC $M^\pi$ that resolves all nondeterminism. The transition probability between any two states $s,s' \in S$ is computed by marginalizing over all actions
\begin{equation}
P^\pi(s,s') = \sum_{a \in Act(s)} \pi(a \mid s) \cdot Tr(s,a,s').
\label{eq:transition}
\end{equation}
We build $M^\pi$ incrementally starting from the initial state $s_0$.
For each reachable state $s$, we consider all actions $a$ with non-zero policy probability ($\pi(a \mid s) > 0$).
For each such action, we expand all successor states $s'$ where $Tr(s,a,s') > 0$ and compute their transition probabilities using Equation~\eqref{eq:transition}.
This process continues until no new states are discovered.
It terminates early once all states relevant to the PCTL property verification have been visited, avoiding unnecessary state expansion that does not affect the verification result.

To assess the impact of a specific quantum noise model, we follow the same procedure in Algorithm~\ref{alg:qverifier} using the noisy policy $\pi_{\theta}^{\mathcal{E}}$ instead of the policy $\pi_{\theta}$. This constructs a separate induced DTMC that captures how hardware imperfections affect the policy's behavior. Figure~\ref{fig:placeholder} illustrates this process for a specific state: starting from state $s_0$, the blue solid arrows show transitions from the noise-free policy, while the red dashed arrows show how the modeled quantum noise alters the transition probabilities in the action sampling process.

\subsection{Verification}
The resulting induced DTMC fully encodes the QRL policy's probabilistic behavior (with and without a quantum noise model), enabling formal verification using the Storm model checker.
Storm verifies whether the specified PCTL property holds and computes exact satisfaction probabilities.

\begin{algorithm}[t]
\caption{QVerifier: Formal Verification of QRL Policies}
\label{alg:qverifier}
\begin{algorithmic}[1]
\Require MDP $M = (S, s_0, Act, Tr, rew, AP, L)$
\Require Trained QRL policy $\pi_{\theta}$
\Require PCTL property $\varphi$
\Ensure Satisfaction result and probability $p$

\Statex
\Statex \textbf{Stage 1: Induced DTMC Construction}
\State $M^\pi \gets \textsc{BuildDTMC}(s_0, \emptyset)$

\Statex
\Statex \textbf{Stage 2: Probabilistic Model Checking}
\State $(result, p) \gets \textsc{Storm}.\text{verify}(M^{\pi}, \varphi)$
\State \Return $(result, p)$

\Statex
\Procedure{BuildDTMC}{$s, Visited$}
    \If{$s \in Visited$ \textbf{or} $s$ is not relevant for $\varphi$}
        \State \Return
    \EndIf
    \State $Visited \gets Visited \cup \{s\}$
    \ForAll{$s' \in S$ where $P^\pi(s, s') > 0$}
        \State Compute $P^\pi(s, s') = \sum_{a \in Act(s)} \pi_{\theta}(a \mid s) \cdot Tr(s, a, s')$
        \State \textsc{BuildDTMC}($s', Visited$)
    \EndFor
\EndProcedure

\end{algorithmic}
\end{algorithm}

\subsection{Limitations}
Our approach supports verifying any memoryless stochastic quantum policy, but it has several constraints.
We are restricted to discrete state and action spaces~\cite{DBLP:conf/setta/GrossJJP22}.
The size of the parameterized quantum circuit also poses scalability limits~\cite{DBLP:journals/corr/abs-2410-12660}. Additionally, time-dependent noise effects such as thermal relaxation~\cite{jang2025balancing} cannot be modeled due to the Markovian property of our formalism~\cite{DBLP:conf/setta/GrossJJP22}.

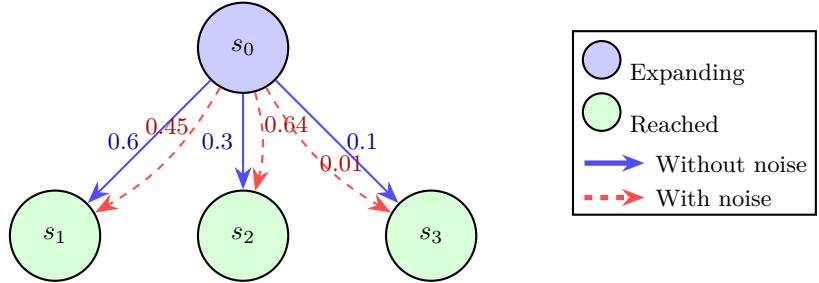
\begin{figure}
    \centering
    \begin{tikzpicture}[
    state/.style={circle, draw, thick, minimum size=1.2cm, font=\normalsize},
    expanded/.style={circle, draw, thick, fill=blue!20, minimum size=1.2cm, font=\normalsize},
    new/.style={circle, draw, thick, fill=green!15, minimum size=1.2cm, font=\normalsize},
    original/.style={-{Stealth[length=3mm]}, thick, blue!70},
    noisy/.style={-{Stealth[length=3mm]}, thick, red!70, dashed},
    >=Stealth
]


\node[expanded] (s0) at (0, 2.5) {$s_0$};

\node[new] (s1) at (-2.5, 0) {$s_1$};
\node[new] (s2) at (0, 0) {$s_2$};
\node[new] (s3) at (2.5, 0) {$s_3$};

\draw[original] (s0) -- (s1) node[midway, left, font=\small, blue!70!black] {0.6};
\draw[original] (s0) -- (s2) node[midway, left, font=\small, blue!70!black] {0.3};
\draw[original] (s0) -- (s3) node[midway, right, font=\small, blue!70!black] {0.1};

\draw[noisy] (s0) to[bend left=15] node[pos=0.35, left, xshift=2mm, yshift=2mm, font=\small, red!70!black] {0.45} (s1);
\draw[noisy] (s0) to[bend left=15] node[pos=0.4, right, xshift=-1mm, yshift=1mm, font=\small, red!70!black] {0.64} (s2);
\draw[noisy] (s0) to[bend right=15] node[pos=0.35, right, xshift=1mm, yshift=-3mm, font=\small, red!70!black] {0.01} (s3);

\node[draw, thick, fill=white, align=left, font=\small] at (6, 1.5) {
    \tikz{\node[expanded, minimum size=0.5cm] {};} Expanding\\[2pt]
    \tikz{\node[new, minimum size=0.5cm] {};} Reached\\[4pt]
    \tikz{\draw[original, line width=2pt] (0,0) -- (0.8,0);} Without noise\\[2pt]
    \tikz{\draw[noisy, line width=2pt] (0,0) -- (0.8,0);} With noise
};
red dashed arrows represent a separate DTMC constructed from the noisy
\end{tikzpicture}
\caption{\emph{One expansion step in the incremental DTMC construction, showing two example DTMCs overlaid for comparison.} Starting from state $s_0$, the policy samples an action which leads to successor states $s_1$, $s_2$, and $s_3$. Blue solid arrows represent the DTMC constructed from the noise-free policy, while red dashed arrows represent a separate DTMC constructed from the noisy policy $\pi_{\theta}^{\mathcal{E}}$, illustrating how the quantum noise model in the action sampling process alters the transition probabilities. This expansion process is repeated for each new state until all states reachable by the policy have been visited.}
    \label{fig:placeholder}
\end{figure}

\section{Experiments}
In this section, we empirically evaluate the verification of QRL policies using probabilistic model checking. Our experiments are designed to address the following research questions:

\begin{itemize}
    \item \textbf{RQ1:} How do quantum and classical policy verification compare?
    \item \textbf{RQ2:} How does quantum noise affect QRL policies?
\end{itemize}

To answer these research questions, we evaluate trained classical and quantum policies across three benchmark environments and verify them against safety properties expressed in PCTL.

We structure our analysis as follows. First, we compare classical REINFORCE with noise-free quantum REINFORCE verification to assess performance differences and potential safety violations.
Second, we incorporate realistic quantum noise models directly into the verification process to study how gate-level noise propagates through the induced stochastic policy and affects key safety-relevant properties.

\subsection{Setup}
We now describe our setup. First, we describe the used environments, then the trained policies, and finally, our technical setup.

\subsubsection{Environments.}
The \emph{Frozen Lake} environment is a commonly used OpenAI gym benchmark, where 
the agent has to reach the goal (frisbee) on a frozen lake (reward +1). The agent's movement direction is partially uncertain.

The \emph{Ski} environment uses a 4-bit state representation encoding the integers 0 to 15, with the agent starting in state 1 and choosing between two actions called left and right. Transitions depend on whether the current state is odd or even: odd states advance through left and even states advance through right, while incorrect choices often send the agent back to state 0, which represents a crash. State 6 serves as the goal and absorbs both actions. Rewards are given when the agent selects the action that correctly advances the current state rather than returning to zero, creating a simple parity-based navigation task toward the goal.

In the \emph{Freeway} environment, the RL agent controls a chicken (up, down, or no operation) running across a highway filled with traffic to get to the other side.
Every time the chicken gets across the highway, it earns a reward of one.
An episode ends if the chicken gets hit by a car or reaches the other side.
Each state is an image of the game's state.
Note that we use an abstraction of the original game, which sets the chicken into the middle column of the screen and contains fewer pixels than the original game, but uses the same reward function and actions~\cite{DBLP:conf/icaart/GrossS24}.

\subsubsection{Trained policies}
In each environment, we train a classical REINFORCE agent and a quantum-enhanced REINFORCE agent with similar training conditions.
Both agents are trained for the same number of epochs (always 10,000 epochs), and we use identical reward structures.
While our focus is on these two types of REINFORCE agents, the approach generalizes to other memoryless QRL policies~\cite{meyer2022survey,jerbi2021parametrized}.
Note that we did not focus on archiving high-performing agents, as this paper focuses on the verification process itself.
For details about the exact training source code, we refer to the technical setup.

\subsubsection{Technical setup}
We executed our benchmarks in a Docker container with 16 GB RAM, and an AMD Ryzen 7 7735hs with Radeon graphics × 16 processor with the operating system Ubuntu 20.04.5 LTS.
For model checking, we use Storm 1.7.1 (dev). For the quantum circuit, we use Pennylane 0.42.3.
Implementation details are provided in the accompanying source code \url{https://github.com/LAVA-LAB/COOL-MC/tree/qverifier}.

\begin{table*}[t]
\centering
\caption{Comparison between QRL and classical RL in probabilistic model checking. The table reports, for each environment and PCTL property, the verification result, the number of explored states and transitions, and the total verification time (building time + model checking time).}
\label{tab:compare}
\scalebox{0.7}{
\begin{tabular}{%
  ll  
  c   
  ccc 
  c   
  cc  
  c   
}
\toprule
\multicolumn{2}{c}{\textbf{Setup}}
  & \multicolumn{4}{c}{\textbf{QRL Policy Model Checking}}
  & \multicolumn{4}{c}{\textbf{RL Policy Model Checking}} \\
\cmidrule(lr){1-2}
\cmidrule(lr){3-6}
\cmidrule(lr){7-10}
\textbf{Env.} & \textbf{PCTL Query}
  & \textbf{Result}
  & \textbf{States} & \textbf{Transitions} & \textbf{Time (s)} 
  & \textbf{Result} & \textbf{States} & \textbf{Transitions} & \textbf{Time (s)} \\
\midrule
Freeway & P($\text{F}$ Goal)
     & 0.621
     & 496 & 2080 & 84
     & 0.7 & 496 & 2080 & 4 \\
\midrule
Frozen Lake & P($\text{F}$ Goal)
     & 0.03
     & 17 & 48 & 6
     & 0.04 &  17 & 48 & 0.2 \\
\midrule
Frozen Lake & $P(pos\leq 3 \text{ U } pos=7)$
     & 0.124
     & 8 & 18 & 1.19
     & 0.061 &  8 & 18 & 0.05 \\
\midrule
Ski & P($\text{F}$ Goal)
     & 0.437
     & 7 & 12 & 0.5
     & 0.45 & 7 & 12 & 0.03 \\

\bottomrule
\end{tabular}
}
\end{table*}

\subsection{Analysis} First, we do a comparative performance analysis between classical REINFORCE and quantum REINFORCE verification. Second, we analyze how different quantum noise models influence the trained quantum REINFORCE policies.
\begin{figure}[t]
    \centering
    \scalebox{0.48}{
        \input{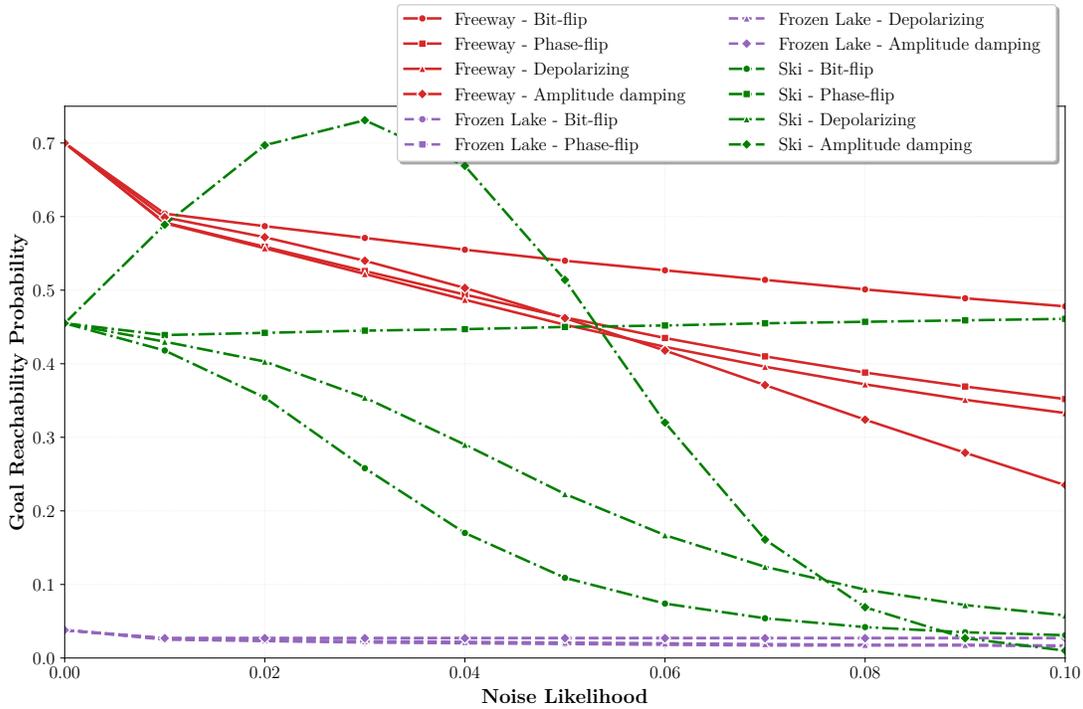}
    }
    \caption{\emph{Performance changes of the quantum policies} under increasing noise likelihoods for the goal-reachability probabilities. 
    The $x$-axis denotes the noise likelihoods, while the $y$-axis indicates the corresponding performance metric (higher is better). }
    \label{fig:placeholder}
\end{figure}

\subsubsection*{RQ1: How do quantum and classical policy verification compare?}
Table~\ref{tab:compare} summarizes the verification outcomes for both classical and quantum policies. For each environment, we report the reachability probability (i.e., the likelihood that the policy reaches the target state) and a more complex until property (i.e., the likelihood that the policy reaches a specific state while satisfying a given condition along the path), the number of states and transitions explored during model checking, and the total verification time.

Across all environments, classical REINFORCE policies outperform their quantum counterparts in terms of reachability probability.
This performance gap may be attributed to the limited expressivity of shallow variational circuits or to optimization challenges inherent in quantum policy gradient methods.
Considering the complex safety requirement, expressed via the until-formula $P(pos\leq 3 \text{ U } pos=7)$, which represents the unsafe behavior of the agent traversing the top border leftward before falling into the water hole at position 7, we find that the QRL policy violates this safety requirement more often than the RL policy.

Quantum policies consistently exhibit longer verification times than classical policies, primarily due to the computational overhead of the quantum circuit at each verification state.

The induced DTMC sizes (states and transitions) are identical for QRL and classical RL policies in each environment, since both assign non-zero probability to all actions, making the same states reachable; only the transition probabilities differ.
However, DTMC sizes can differ when certain actions have zero probability.

These results suggest that, under ideal (noise-free) conditions, classical policies remain more effective for the tasks considered.
However, as discussed in the following section, introducing realistic noise models can alter this comparison, sometimes leading to scenarios in which quantum policies outperform classical policies.

\subsubsection*{RQ2: How does quantum noise affect QRL policies?}
To evaluate how quantum hardware imperfections affect policy reliability, we conduct experiments using a gate-level noise model integrated directly into the policy verification procedures, focusing on the goal-reachability probabilities across the different environments.

In our experiment, at \textit{every} encountered state $s$, the agent computes its action distribution by executing the full noisy quantum circuit.
Concretely, the process consists of:  
(i) encoding the classical state $s$ into the quantum amplitudes,  
(ii) executing the variational quantum circuit with noise inserted after each gate, and  
(iii) obtaining the stochastic policy $\pi_{\theta}^{\mathcal{E}}$.  
In this setup, noise accumulates proportionally to the circuit depth, so deeper circuits experience greater~degradation.

Figure~\ref{fig:placeholder} shows how performance changes across environments as the probability of applying noise at each gate increases. We study several standard noise channels: bit-flip, phase-flip, depolarizing, and amplitude damping. The verification results vary across noise types.

Bit-flip noise inverts computational basis states, causing performance loss across all tasks.

Depolarizing noise introduces a uniform mixture of Pauli errors, pushing quantum states toward maximal randomness and resulting in the strongest overall degradation.

Phase-flip noise, which affects only relative phases while leaving populations unchanged, shows a slight performance improvement for the QRL Ski policy.

Finally, we observe an effect under low levels of amplitude-damping noise: in the Ski environment, performance increases and even surpasses the trained classical RL baseline by 27\% (see Table~\ref{tab:compare} for comparison).
This beneficial regime disappears as noise levels increase, suggesting that mild dissipation can occasionally stabilize or regularize quantum policies, whereas stronger noise ultimately degrades performance.
These observations align with prior findings~\cite{jerbi2021quantum}, which similarly report noise-induced performance enhancements.

\section{Discussion}
\emph{QVerifier} addresses the high cost and limited availability of quantum hardware~\cite{maring2023one,DBLP:journals/sqj/MoguelRVBGM22}: verifying QRL policies classically avoids the expense of on-device testing~\cite{DBLP:journals/sqj/MoguelRVBGM22}.
QRL policy verification reveals potential safety violations, shows how a specific quantum noise model affects policy behavior, and supports deployment decisions.

A key feature is the \emph{exact} analytical treatment of quantum noise.
Instead of relying on empirical sampling, we compute action probabilities directly from the density matrix, yielding noise-dependent exact transition probabilities.
This enables continuous noise-parameter sweeps and precise analysis of how safety properties change with varying noise likelihoods~(see Figure~\ref{fig:placeholder}).
When the policy was trained on a simulator with matching noise conditions, the resulting formal model is exact.
When trained on physical hardware, however, the model constitutes an idealized approximation, as unknown hardware noise prevents exact modeling.
Nevertheless, even idealized models provide valuable insights: they reveal how policies respond to specific noise types and magnitudes, enabling comparative analysis and identification of fragile designs.

The resulting noise–safety profiles guide hardware selection~\cite{phalak2025qualiti}. If, for example, a policy remains safe up to a depolarizing rate of $p=0.05$, practitioners can choose devices whose gate fidelities meet this requirement and allocate error budgets accordingly.

When full environment verification is infeasible~\cite{DBLP:conf/setta/GrossJJP22}, \emph{QVerifier} still supports targeted debugging.
Practitioners can analyze critical regions of the state space to locate unsafe behaviors.

A fundamental tension underlies any classical verification of quantum systems~\cite{babbush2025grand,meyer2022survey,bowles2024better}: if a circuit is small enough to compute efficiently, one might question the benefit of quantum execution; if it is too large, classical verification becomes too costly~\cite{DBLP:journals/corr/abs-2410-12660}.
\emph{QVerifier} targets a \emph{sweet spot}: trained QRL policies that run efficiently on quantum hardware and are still computable on classical hardware for model checking.
This enables \emph{QVerifier} to analyze different QRL policy designs~\cite{meyer2022survey}, allowing systematic comparisons~\cite{bowles2024better} of how architectural choices, such as circuit depth~\cite{fosel2021quantum}, encoding schemes~\cite{rath2024quantum}, or measurement strategies~\cite{li2001quantum}, impact robustness under noise~\cite{DBLP:journals/corr/abs-2410-21117} and adherence to safety requirements, including for smaller quantum circuits.
Note that \emph{QVerifier} can also be used without quantum noise, which is particularly relevant for quantum-inspired reinforcement learning~\cite{habibi2025quantum} approaches that employ quantum formalizations in their classical policy design without requiring quantum hardware.

Beyond its immediate applications, \emph{QVerifier} may inform the design of future quantum model checkers~\cite{baltazar2008quantum}.

\section{Conclusion}
We proposed \emph{QVerifier}, a method for formally verifying trained QRL policies against safety properties while accounting for quantum measurement uncertainty and quantum noise.
We validated our method across multiple scenarios with and without quantum noise models.

Future work includes integrating explainable RL methods to attribute safety performance to specific gates and circuit elements~\cite{DBLP:conf/esann/GrossS24}, using model-checking feedback to guide training so that QRL agents learn better policies~\cite{gu2024review}, and quantum noise mitigation~\cite{muqeet2025quiet} for QRL.

\section*{Acknowledgements}
I thank \emph{Shaukat Ali} for his valuable feedback, his expertise, and his time. His guidance helped shape several key ideas and greatly improved the clarity and quality of the final presentation.

\bibliographystyle{plain}
\bibliography{refs}

\end{document}